\documentclass{PoS}
\usepackage{epsf,amssymb,amsthm,amsfonts,amsmath,latexsym, graphicx,array,extarrows,mathtools,mathstyle,mathrsfs,slashed}

\newcommand{\overbar}[1]{\mkern 1.5mu\overline{\mkern-1.5mu#1\mkern-1.5mu}\mkern 1.5mu}

\title{On the analytic structure of QCD propagators}

\ShortTitle{On the analytic structure of QCD propagators}

\author{\speaker{Peter Lowdon}\\
        SLAC National Accelerator Laboratory, Stanford University, USA\\
        E-mail: \email{lowdon@slac.stanford.edu}}

\abstract{Local formulations of quantum field theory provide a powerful framework in which non-perturbative aspects of QCD can be analysed. Here we report on how this approach can be used to elucidate the general analytic features of QCD propagators, and why this is relevant for understanding confinement.}

\FullConference{XIII Quark Confinement and the Hadron Spectrum - Confinement2018\\
		31 July - 6 August 2018\\
		Maynooth University, Ireland}

\begin{document}

\section{Introduction}

\noindent
The non-perturbative behaviour of propagators involving coloured fields plays an important role in many areas of quantum chromodynamics (QCD), from the dynamics of quark-gluon plasma to the nature of confinement itself~\cite{Kugo_Ojima79,Nakanishi_Ojima90,Alkofer_vonSmekal01, Alkofer_Greensite07,Lowdon16,Lowdon17_1}. Nevertheless, the overall analytic structure of these objects remains largely unknown. In order to gain a better understanding of the general characteristics of these objects one necessarily requires a non-perturbative approach. An important such example are local formulations of quantum field theory (QFT), the construction of which are based on a series of physically motivated axioms~\cite{LQFT}. A significant advantage of this framework is that the axioms are assumed to hold independently of the coupling regime, allowing non-perturbative features to be derived in a purely analytic manner. Numerical non-perturbative techniques such as lattice Monte-Carlo simulations and functional methods have also played an essential role in helping to unravel the structure of QCD propagators, especially in recent years~\cite{Alkofer_vonSmekal01,Alkofer_Detmold_Fischer_Maris04,Cucchieri_Mendes_Taurines05,Cucchieri_Mendes08,Oliveira_Silva09,Strauss_Fischer_Kellermann12, Dudal_Oliveira_Silva14}. However, significant progress has been achieved when local QFT has been used both as a guide for the analytic input required for these numerical techniques, and also to help interpret the corresponding results. Here we report on recent progress~\cite{Lowdon17_1,Lowdon17_2,Lowdon18} in establishing the structural form of QCD propagators using a local QFT approach.

\section{The general structure of correlators in local QFT}

\noindent
In local formulations of QFT a characteristic of central importance is that correlators are \textit{distributions}~\cite{LQFT}. Distributions are a generalisation of functions, and among other things this implies that they can possess a broader range analytic properties compared to regular functions. Due to the Lorentz transformation properties of the fields $\phi_{1}$ and $\phi_{2}$ it follows that the Fourier transform of any correlator $\widehat{T}_{(1,2)}(p) = \mathcal{F}\left[\langle 0 | \phi_{1}(x)\phi_{2}(y) |0\rangle\right]$ has the decomposition
\begin{align}
\widehat{T}_{(1,2)}(p) = \sum_{\alpha=1}^{\mathscr{N}}Q_{\alpha}(p) \, \widehat{T}^{\alpha}_{(1,2)}(p)
\label{decomp_cov}
\end{align}    
where $Q_{\alpha}(p)$ are Lorentz covariant polynomial functions\footnote{For example, when $\phi_{1}=\psi$ and $\phi_{2}=\overbar{\psi}$ there are two such functions: $Q_{1}(p) = \mathbb{I}$ and $Q_{2}(p) = \slashed{p}$.} of $p$ carrying the same Lorentz index structure as $\phi_{1}$ and $\phi_{2}$, and $\widehat{T}^{\alpha}_{(1,2)}(p)$ are \textit{Lorentz invariant} distributions\footnote{Lorentz invariant distributions satisfy the property $\widehat{T}^{\alpha}_{(1,2)}(\Lambda p)= \widehat{T}^{\alpha}_{(1,2)}(p)$ for any Lorentz transformation $\Lambda$.}~\cite{Bogolubov_Logunov_Oksak90}. Due to the decomposition in Eq.~(\ref{decomp_cov}), the key to defining any correlator is to understand the properties of the Lorentz invariant distributional components. In principle these objects could have a wide variety of different properties, but it turns out that the physical requirement for states in the theory to have positive energy implies that the components $\widehat{T}^{\alpha}_{(1,2)}(p)$ must \textit{vanish} outside the closed forward light cone $\overbar{V}^{+}= \{p^{\mu} \ | \ p^{2} \geq 0, \, p^{0}\geq 0 \}$, and can therefore be written in the following general manner~\cite{Bogolubov_Logunov_Oksak90}:   
\begin{align}
\widehat{T}^{\alpha}_{(1,2)}(p) =  \int_{0}^{\infty} ds \, \theta(p^{0})\delta(p^{2}-s) \rho_{\alpha}(s) + P_{\alpha}(\partial^{2})\delta(p) 
\label{KL_gen_rep} 
\end{align}   
where $P_{\alpha}(\partial^{2})$ is a polynomial of finite order in the d'Alembert operator $\partial^{2} = g_{\mu\nu}\frac{\partial}{\partial p_{\mu}}\frac{\partial}{\partial p_{\nu}}$, and $\rho_{\alpha}(s)$ are distributions with support in $\overbar{\mathbb{R}}_{+}$. Eq.~(\ref{KL_gen_rep}) is the so-called spectral representation of $\widehat{T}^{\alpha}_{(1,2)}(p)$, and $\rho_{\alpha}(s)$ the spectral density. If one instead considers the time-ordered correlator (propagator) $\widehat{D}_{(1,2)}(p) = \mathcal{F}\left[\theta(x^{0}-y^{0})\langle 0 | \phi_{1}(x)\phi_{2}(y) |0\rangle   + \sigma_{(1,2)}\theta(y^{0}-x^{0}) \langle 0 | \phi_{2}(y)\phi_{1}(x) |0\rangle \right]$, where $\sigma_{(1,2)}= \pm 1$ depending on the spin statistics of the fields, it follows from Eq.~(\ref{KL_gen_rep}) that the corresponding Lorentz invariant components $\widehat{D}^{\alpha}_{(1,2)}(p)$ have the structure  
\begin{align}
\widehat{D}^{\alpha}_{(1,2)}(p) = i \int ds \, \frac{\rho_{\alpha}(s)}{p^{2}-s+i\epsilon} + P_{\alpha}(\partial^{2})\delta(p)
\label{spec_rep}
\end{align}
The first term has the familiar looking K\"{a}ll\'{e}n-Lehmann spectral form, whereas the second term is purely singular. In theories for which the space of states have a positive-definite inner product it turns out that $P_{\alpha}(\partial^{2})\delta(p)$ can only contain terms proportional to $\delta(p)$~\cite{Bogolubov_Logunov_Oksak90}. An important feature of gauge theories such as QCD is that the gauge symmetry provides an obstacle to the locality of the theory~\cite{Strocchi13}. In order to consistently quantise the theory one is therefore left with two options: explicitly preserve locality, or allow non-local fields. A general feature of local quantisations is that additional degrees of freedom are introduced into the theory, resulting in a space of states with an indefinite inner product. The prototypical example is the Becchi-Rouet-Stora-Tyutin (BRST) quantisation of QCD, where the space of states contains negative-norm ghost states~\cite{Nakanishi_Ojima90}. Although many features of positive-definite inner product QFTs are preserved in BRST quantised QCD, the existence of an indefinite inner product can lead to significant changes to the structure of propagators. In particular, $P_{\alpha}(\partial^{2})\delta(p)$ can potentially contain terms involving derivatives of $\delta(p)$~\cite{Lowdon17_1}. The relevance of these types of contribution was first recognised in~\cite{Strocchi76,Strocchi78}, where it was proven that their presence can fundamentally alter the asymptotic behaviour of correlators, and in fact cause the correlation strength between clusters of states to grow with distance. For clusters of coloured states this provides a mechanism which can guarantee their absence from the asymptotic spectrum, since a growth in correlation strength between coloured states prevents the independent measurement of either of these states at large distances. In other words, the presence of these type of contributions are indicative of confinement~\cite{Lowdon16}. 

\section{Dynamical constraints on the QCD propagators}

\noindent
In light of the general structural features of propagators in locally quantised QCD, determining the behaviour of propagators associated with coloured fields is important for understanding the non-perturbative dynamics of the theory, and in particular confinement. Since the quark, gluon and ghost fields parametrise the coloured degrees of freedom in this theory, the propagators associated with these fields play a crucial role. With this motivation in mind, in Refs.~\cite{Lowdon17_1,Lowdon17_2,Lowdon18} a local QFT approach was adopted in order to derive the most general structural form of these QCD propagators, and the constraints imposed on them by the dynamical properties of the theory\footnote{For an alternative approach see e.g.~\cite{Lowdon15}.}. \\

\noindent
In the case of the quark propagator $\widehat{S}_{F}^{ij}(p)= \mathcal{F}\left[\langle 0|T\{ \psi^{i}(x)\overbar{\psi}^{j}(y)\}|0\rangle\right]$ it follows from Eqs.~(\ref{KL_gen_rep}) and~(\ref{spec_rep}) that the propagator can be written~\cite{Lowdon17_2} 
\begin{align}
\widehat{S}_{F}^{ij}(p) &=   i\int_{0}^{\infty} \frac{ds}{2\pi} \, \frac{\left[ \rho_{1}^{ij}(s) + \slashed{p}\rho_{2}^{ij}(s) \right]}{p^{2}-s +i\epsilon}  + \left[ P_{1}^{ij}(\partial^{2}) + \slashed{p} P_{2}^{ij}(\partial^{2})\right]\delta(p) 
\label{quark_propagator_mom}
\end{align} 
It turns out that the equations of motion impose constraints both on the form of the spectral densities, and the coefficients of the potential singular terms. In Ref.~\cite{Lowdon17_2} it was demonstrated that these constraints can be derived by considering the Dyson-Schwinger equation, which in momentum space has the form   
\begin{align}
(\slashed{p} - m)\widehat{S}^{ij}_{F}(p) = i\delta^{ij}Z_{2}^{-1} + \widehat{K}^{ij}(p)
\label{quark_sde_p}
\end{align}
where $Z_{2}$ is the quark field renormalisation constant, and $\widehat{K}^{ij}(p)$ is the current-quark propagator. By inserting the spectral representation of $\widehat{S}_{F}^{ij}(p)$ and the corresponding representation of $\widehat{K}^{ij}(p)$ one can match the different Lorentz components on both sides, and this give rise to a series of constraints. After applying this procedure one finds that the coefficients of the singular terms in the propagators are linearly related to one another, and that the quark spectral densities have the following representation~\cite{Lowdon17_2}
\begin{align}
&\rho_{1}^{ij}(s) = \left[2\pi m \, \delta^{ij} Z_{2}^{-1} - \int d\tilde{s} \, \kappa_{1}^{ij}(\tilde{s})    \right] \delta(s-m^{2}) + \kappa_{1}^{ij}(s) \\
&\rho_{2}^{ij}(s) =\left[2\pi \delta^{ij} Z_{2}^{-1} - \int d\tilde{s} \, \kappa_{2}^{ij}(\tilde{s})   \right]  \delta(s-m^{2}) + \kappa_{2}^{ij}(s)
\end{align}
These equalities explicitly demonstrate that both spectral densities contain a discrete mass component, but that the coefficients in front of these components depend explicitly on the behaviour of $\kappa_{1}^{ij}(s)$ and $\kappa_{2}^{ij}(s)$, both of which are related to the corresponding spectral densities of the current-quark propagator $\widehat{K}^{ij}(p)$.  \\

\noindent
An analogous approach as applied to the quark propagator can also be used to constrain the ghost propagator $\widehat{G}_{F}^{ab}(p) = \mathcal{F}\left[\langle 0|T\{ C^{a}(x)\overbar{C}^{b}(y)\}|0\rangle\right]$~\cite{Lowdon17_2}. In this case the propagator has the general form
\begin{align}
\widehat{G}_{F}^{ab}(p) =   i\int_{0}^{\infty} \frac{ds}{2\pi} \, \frac{ \rho_{C}^{ab}(s) }{p^{2}-s +i\epsilon}  + P_{C}^{ab}(\partial^{2})\delta(p)
\label{ghost_prop_mom_QCD}
\end{align}
and the momentum space Dyson-Schwinger equation is given by
\begin{align}
-p^{2}  G^{ab}_{F}(p) = \delta^{ab}\widetilde{Z}_{3}^{-1} + L^{ab}(p)
\label{gh_sde_p}
\end{align} 
where now $\widetilde{Z}_{3}$ is the ghost field renormalisation constant, and $L^{ab}(p)$ is the current-ghost propagator. Inserting the spectral representations of these propagators into this equation one again finds that the coefficients of the singular terms in both propagators are linearly related to one another, and that the ghost spectral density is constrained to satisfy~\cite{Lowdon17_2}:   
\begin{align}
\rho_{C}^{ab}(s) = \left[2\pi i\delta^{ab}\widetilde{Z}_{3}^{-1} - \int_{0}^{\infty} d\tilde{s} \, \kappa_{C}^{ab}(\tilde{s})   \right] \delta(s) + \kappa_{C}^{ab}(s)
\label{ghost_rho}
\end{align}
Eq.~(\ref{ghost_rho}) demonstrates that the ghost spectral density contains a discrete massless component. However, similarly to the quark spectral densities, the coefficient in front of this component is not completely constrained since it depends on the integral of $\kappa_{C}^{ab}(s)$, which itself is determined by the spectral function of $L^{ab}(p)$. Therefore, the presence or absence of a non-perturbative massless ghost pole is not so clear-cut. \\

\noindent
The final QCD propagator of interest involves the gluon field. In this case the propagator has the general form~\cite{Lowdon17_1} 
\begin{align}
\widehat{D}_{\mu\nu}^{ab\, F}(p) &=   i\int_{0}^{\infty} \frac{ds}{2\pi} \, \frac{\left[ g_{\mu\nu}\rho_{1}^{ab}(s) + p_{\mu}p_{\nu}\rho_{2}^{ab}(s) \right]}{p^{2}-s +i\epsilon}   +\sum_{n=0}^{N} \left[ c_{n}^{ab} \, g_{\mu\nu} (\partial^{2})^{n} + d_{n}^{ab} \partial_{\mu}\partial_{\nu}(\partial^{2})^{n-1}\right]\delta(p)
\label{general_propagator_QCD_mom}
\end{align}
The Dyson-Schwinger equation for this propagator is given by
\begin{align}
-\left[ p^{2}g_{\mu}^{\ \alpha} - \left(1 - \frac{1}{\xi_{0}} \right)p_{\mu}p^{\alpha}  \right]\widehat{D}_{\alpha\nu}^{ab\, F}(p) = i\delta^{ab} g_{\mu  \nu}Z_{3}^{-1} + \widehat{J}_{\mu\nu}^{ab}(p)
\label{SDE_p_gluon}
\end{align}
where $Z_{3}$ is the gluon field renormalisation constant, $\widehat{J}_{\mu\nu}^{ab}(p)$ the current-gluon propagator, and $\xi_{0}$ is the bare gauge fixing parameter. Again, by inserting the spectral representations of the gluon and current-gluon propagators one obtains constraints. Similarly to the quark and ghost propagators, Eq.~(\ref{SDE_p_gluon}) implies that the coefficients of the potential singular terms in the gluon propagator are linearly related to those in $\widehat{J}_{\mu\nu}^{ab}(p)$. Moreover, the gluon spectral densities are constrained to satisfy the relations
\begin{align}
&\rho_{1}^{ab}(s) + s \rho_{2}^{ab}(s) = -2\pi\xi\delta^{ab}\delta(s)  \\
&\rho_{1}^{ab}(s) = -2\pi \delta^{ab}Z_{3}^{-1}\delta(s) + \widetilde{\rho}_{2}^{ab}(s)  \label{constr_rho1} \\
&\int_{0}^{\infty}ds \, \widetilde{\rho}_{2}^{ab}(s) = 0 \label{constr_rho3_2} 
\end{align}
In contrast to both the quark and ghost spectral densities, Eq.~(\ref{constr_rho1}) implies that $\rho_{1}^{ab}(s)$ contains an explicit massless contribution, and that the coefficient of this component is completely specified by the value of the corresponding renormalisation constant. Since $Z_{3}^{-1}$ is expected to vanish in Landau gauge~\cite{Alkofer_vonSmekal01}, this implies that massless gluons must therefore necessarily be absent from the spectrum in this gauge. In the literature~\cite{Alkofer_Detmold_Fischer_Maris04,Cucchieri_Mendes_Taurines05, Strauss_Fischer_Kellermann12,Dudal_Oliveira_Silva14,Cornwall13} it is often argued that the violation of non-negativity of $\rho_{1}^{ab}(s)$ in Landau gauge as a result of the sum rule\footnote{This sum rule is often referred to as the Oehme-Zimmermann \textit{superconvergence relation}~\cite{Oehme_Zimmermann80_1,Oehme_Zimmermann80_2}.}: $\int ds \, \rho_{1}^{ab}(s)=0$ is the reason why gluons do not appear in the spectrum. However, from the structure of Eq.~(\ref{constr_rho1}) it is apparent that (continuous) non-negativity violations can only arise from the component $\widetilde{\rho}_{2}^{ab}(s)$, which has vanishing integral [Eq.~(\ref{constr_rho3_2})]. Performing an identical analysis for the photon propagator it turns out that this propagator satisfies identical constraints, and in particular the analogous component $\widetilde{\rho}_{2}(s)$ of the photon spectral density $\rho_{1}(s)$ has vanishing integral. This implies that potential non-negativity violations are not QCD specific, and casts doubt on the hypothesis that these violations in Landau gauge are the reason why gluons are absent from the spectrum.

\section{Conclusions}

\noindent
Although the propagators in QCD play an important role in determining the non-perturbative characteristics of the theory, the analytic behaviour of these objects remains largely unknown. It turns out that the dynamical properties of the quark, ghost and gluon fields, and in particular their corresponding Dyson-Schwinger equations, impose considerable constraints on the structure of these propagators. In all of these cases singular terms involving derivatives of $\delta(p)$ are permitted, which is particularly interesting in the context of confinement, and the general form of the corresponding spectral densities are constrained. Besides the purely theoretical relevance of these results, these constraints could also provide important input for improving existing parametrisations of the propagators.

\section*{Acknowledgements}
\noindent
This work was supported by the Swiss National Science Foundation under contract P2ZHP2\_168622, and by the DOE under contract DE-AC02-76SF00515.

\end{document}